\documentclass[]{spie}  %>>> use for US letter paper
\usepackage[]{graphicx}
\usepackage{fancyvrb} % for verbatim stuff
\usepackage{float}
\usepackage{savesym}
\usepackage{listings}
\savesymbol{tablenum}
\usepackage{siunitx}
\restoresymbol{SIX}{tablenum}
\usepackage{subfigure}

\usepackage{color}
\newcommand\crispy{\Verb|crispy|}

\DeclareSIUnit\elec{\ensuremath{e^-}}
\DeclareSIUnit\frame{frame}
\DeclareSIUnit\pixel{pixel}

\title{Simulating the WFIRST coronagraph Integral Field Spectrograph} 

\author{Maxime J. Rizzo\supit{a,e}, Tyler D. Groff\supit{a}, Neil T. Zimmerman\supit{a},  Qian Gong\supit{a}, Avi M. Mandell\supit{a}, Prabal Saxena\supit{a,e}, Michael W. McElwain\supit{a}, Aki Roberge\supit{a}, John~Krist\supit{b}, A~J~Eldorado~Riggs\supit{b},
Eric~J.~Cady\supit{b}, Camilo~Mejia~Prada\supit{b}, Timothy Brandt\supit{c}, Ewan Douglas\supit{d}, Kerri Cahoy\supit{d}
\skiplinehalf
\supit{a}NASA Goddard Space Flight Center, 8800 Greenbelt Road, Greenbelt, MD, 20771, USA \\
\supit{b}Jet Propulsion Laboratory, California Institute of Technology, Pasadena, CA, USA\\
\supit{c}Institute for Advanced Studies, Princeton, NJ, USA\\
\supit{d}Massachussetts Institute of Technology, Cambridge, MA, USA\\
\supit{e}NASA Postdoctoral Fellow
}
\authorinfo{Send correspondence to: maxime.j.rizzo@nasa.gov}
\begin{document} 
  \maketitle 

%%%%%%%%%%%%%%%%%%%%%%%%%%%%%%%%%%%%%%%%%%%%%%%%%%%%%%%%%%%%% 
\begin{abstract}
A primary goal of direct imaging techniques is to spectrally characterize the atmospheres of planets around other stars at extremely high contrast levels. To achieve this goal, coronagraphic instruments have favored integral field spectrographs (IFS) as the science cameras to disperse the entire search area at
once and obtain spectra at each location, since the planet position is not known a priori. These spectrographs are useful against confusion from speckles and background objects, and can also help in the speckle subtraction and wavefront control stages of the coronagraphic observation. We present a software package, the Coronagraph and Rapid Imaging Spectrograph in Python (\crispy) to simulate the IFS of the WFIRST Coronagraph Instrument (CGI). The software propagates input science cubes using spatially and spectrally resolved coronagraphic focal plane cubes, transforms them into IFS detector maps and ultimately reconstructs the spatio-spectral input scene as a 3D datacube. Simulated IFS cubes can be used to test data extraction techniques, refine sensitivity analyses and carry out design trade studies of the flight CGI-IFS instrument. \crispy{} is a publicly available Python package and can be adapted to other IFS designs. \end{abstract}

%>>>> Include a list of keywords after the abstract 


%%%%%%%%%%%%%%%%%%%%%%%%%%%%%%%%%%%%%%%%%%%%%%%%%%%%%%%%%%%%%
\section{INTRODUCTION\label{sec:Introduction}}
 
Recent progress in the maturation of high-contrast techniques have put the goal of directly imaging extrasolar planets from space within reach. WFIRST is the first space observatory to baseline an active coronagraph for high-contrast imaging of exoplanets. It leverages the extreme stability of the space environment and the lack of atmospheric perturbations to reach raw contrasts levels orders of magnitude deeper than what can be achieved from the ground. This allows us to observe a class of planets which have eluded researchers until now: cold, giant exoplanets. Although a number of these planets have been detected through radial velocity techniques, they are too dim to be directly imaged at any wavelength with the state-of-the-art ground-based facilities. Characterizing the composition of this family of objects is a key step towards learning about systems which could harbor habitable, rocky planets, and will pave the way for missions with the ability to directly image exoEarths, like LUVOIR or HabEx.

The WFIRST Coronagraphic Instrument (CGI) will operate in several bands between 400 and \SI{970}{\nano\meter}, observing the starlight that scatters off of the planet's atmosphere. In addition to being scientifically compelling, this wavelength range also leverages recent progress in photon counting detectors which offer the best possible noise characteristics to date\cite{Harding:2016bc}, as well as progress in coronagraphic mask designs which allow dark holes with bandpasses as wide as $\Delta\lambda/\lambda = 18\%$\cite{Zimmerman:2015js}. In order to reach extreme contrasts, CGI features two deformable mirrors operated in closed loop, with data measured directly at the focal plane, and using techniques such as electric field conjugation\cite{GiveOn:2009ey}.

CGI will also feature a lenslet array-based Integral Field Spectrograph (IFS)\cite{Mandell:2017} as a backend instrument to the coronagraph. This instrument is planned to operate in three 18\% bands between 600 and \SI{970}{\nano\meter}. The IFS yields information on the spectral content of the source over the entire band without having to cycle through individual narrow-band filters. This allows us to extract spectra at each point within the entire field of view and presents two fundamental advantages: first, for the locations in the field of view where our planetary targets are located, we can obtain reflected light spectra with a modest spectral resolution over a wide band, hence measuring atmospheric absorption features and determining elemental abundances. Second, the information can be used during operations to actually reach the desired contrasts in the dark hole faster than traditional methods, as both the spatial and spectral dependence of the speckle field are measured at once. Until this method was successfully demonstrated by the PISCES instrument\cite{Saxena:2017,McElwain:2016cb} at JPL's High Contrast Imaging Testbed\cite{Groff:2017}, the standard approach to reach broadband contrast was to cycle through narrow band filters spread over the band, which is costly both in terms of filter slots and in terms of time on target.

Because CGI requires a closed loop to successfully reach deep contrast and observe its targets, an extensive amount of work has been achieved to accurately simulate CGI and its components. This is critical to determine the sensitivity of the instrument to perturbations, the tolerances of its elements, and its overall on-sky science yield. In addition, because coronagraphic data products typically require a significant amount of post-processing to reach the contrasts that the science requires, it is difficult to assess the scientific yield of the instrument without properly modelling its data products. For this simulation effort, we need to simulate an actual CGI observation scenario as a time series of focal plane maps, which include perturbations from the telescope and coronagraph as well as detector noise over the entire duration of the observation. Until recently, this effort has largely been carried out by JPL\cite{Krist:2016gz} and was mostly focused on the Hybrid Lyot Coronagraph (HLC) design which is one of the two types of coronagraphs that is baselined for CGI. The IFS is currently baselined to be used exclusively with the Shaped Pupil Coronagraph (SPC). 

In this paper, we discuss our work in extending the simulation efforts to the SPC and IFS. We use the existing simulation framework to simulate spatially and spectrally resolved focal plane time series, and propagate them through our own parametric IFS instrument simulator in order to generate noisy IFS detector maps. Because in an IFS the spectral and spatial information are encoded onto a single IFS detector map, we also need an algorithm to extract the 2D map back into a spatio-spectral cube. These cubes are the elemental IFS data products that can be made available to scientists during science operations of the observatory. By producing realistic time series of IFS data products, we can identify post-processing methods, observing scenarios, and instrumental parameters that optimize the science yield. 

We organize our paper in two sections. Section~\ref{sec:SoftwareDescription} describes the open-source software package that we wrote to simulate the IFS and decode its detector plane. Section~\ref{sec:Application to WFIRST CGI-IFS} discusses how we applied this software to the CGI-IFS, including a fiducial observing scenario and the results of a trade-off on the lenslet array sampling.

\section{Software description}\label{sec:SoftwareDescription}

We develop the Coronagraph and Rapid Imaging Spectrograph in Python (\crispy)\footnote{{\tt https://github.com/mjrfringes/crispy}}, an open-source software package which can produce end-to-end IFS simulations. It consists of three main parts: the IFS propagation from the lenslet plane to the IFS detector plane (Section~\ref{subsec:IFSPropagation}), the detector model (Section~\ref{subsec:StochasticEMCCDModel}), and the IFS extraction (Section~\ref{subsec:IFSExtraction}). The software is constructed in the context of the WFIRST IFS, but is adaptable to all lenslet array-based IFS designs. The extraction module of \crispy{} is also routinely used for PISCES operations at HCIT\cite{Groff:2017}. Before tackling the software, we first give a brief overview of the IFS instrument design.

\subsection{IFS instrument design}\label{subsec:GeneralLayout}

\begin{figure}[ht!]
\centering
\includegraphics[width=\textwidth]{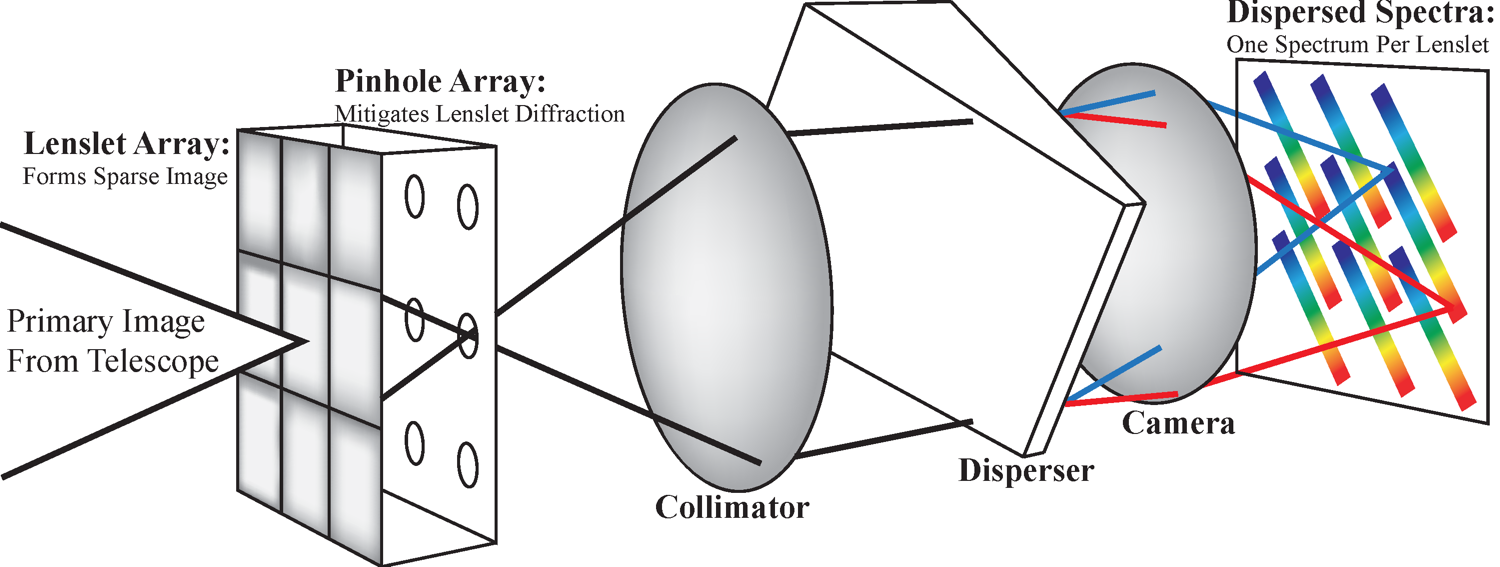}
\caption{Layout of an lenslet array-based IFS. The focal plan of the coronagraph is sampled by the lenslet array. Each lenslet focuses a portion of the focal plane onto a pinhole (for stray light control and cross-talk considerations). The pinhole array plane is then collimated, dispersed with a prism and re-imaged onto the IFS detector array. Each lenslet, which corresponds to a small section of the coronagraphic image plane, now has its own micro-spectrum on the detector and a spatio-spectral cube can be reconstructed. The prism and lenslet array are clocked to avoid overlapping the microspectra and optimize the detector coverage.}\label{fig:IFS}
\end{figure}

%\subsubsection{IFS principle and layout}\label{subsec:IFSPrincipleAndLayout}
The lenslet array-based IFS design and concept is presented in the PISCES publications\cite{Saxena:2017,McElwain:2016cb}. An overall layout is repeated in Fig.~\ref{fig:IFS}. The CGI focal plane is sampled by a square lenslet array, to the left of Fig.~\ref{fig:IFS}. Each square lenslet focuses onto a pinhole array to limit crosstalk between lenslets. The entire field, composed of many spots corresponding to each lenslet/pinhole pair, is then dispersed by a prism and re-imaged onto a detector. Each lenslet now has a corresponding micro-spectrum on the final detector. By choosing a proper clocking angle of the lenslet array with respect to the prism and detector, the microspectra can be aligned in order not to overlap with their neighbors. In the limit where the focal plane at the lenslet array is sampled at the telescope's Nyquist spatial frequency for a given wavelength, and sampled at the Nyquist spectral frequency by the prism/detector, the field at the IFS detector contains the same amount of information as the incident focal plane, but the light has been re-arranged. 

\begin{figure}[ht!]
\centering
\includegraphics[width=\textwidth]{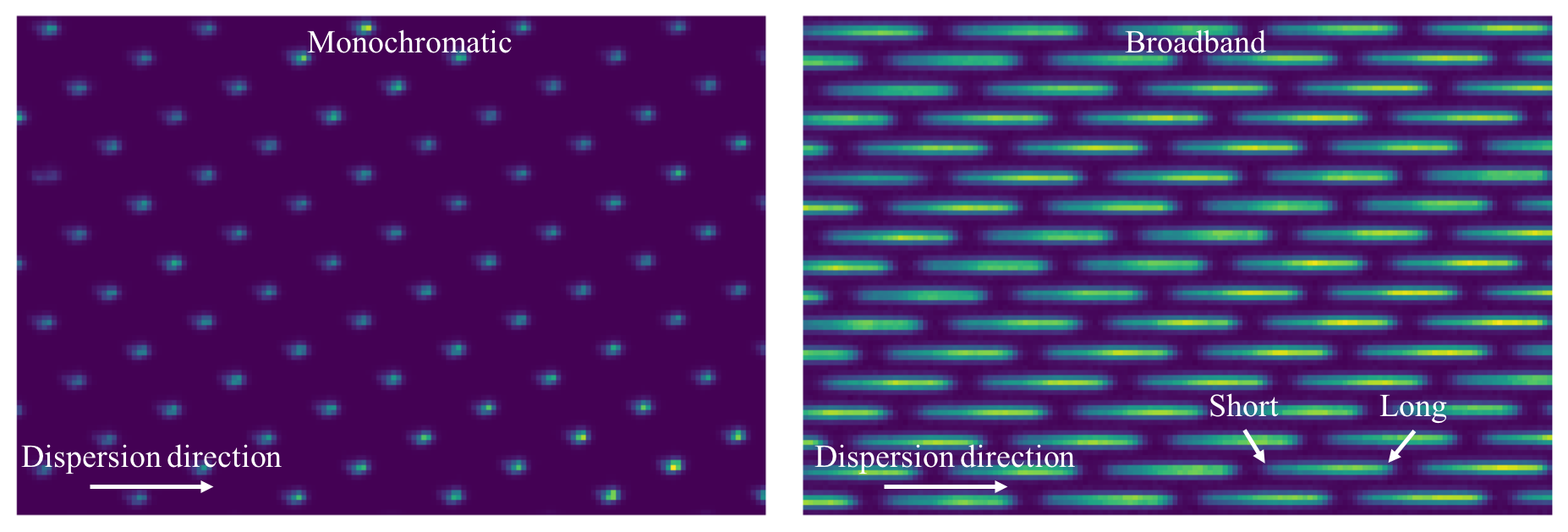}
\caption{Left: Monochromatic grid of spots on the detector. This image is real data from the PISCES testbed and corresponds to a quasi-uniformly illuminated lenslet array in monochromatic light at 637~nm. Right: broadband microspectra grid from PISCES, corresponding to a quasi-uniformly illuminated lenslet array with broad spectrum from 600 to 720~nm. In both images, the dispersion direction is horizontal. The edges of the microspectrum correspond to the short and long end of the band.}\label{fig:flatBB}
\end{figure}

Fig.~\ref{fig:flatBB} shows example detector maps of the IFS in the case of monochromatic and broadband flatfields. In the monochromatic image, the grid of spots represents light from all lenslet/pinhole pairs. Each monochromatic spot is called a PSFLet. In the broadband case, each lenslet has its own microspectrum. While the inherent monochromatic PSFLets are similar for neighboring lenslets, they are sampled differently by the detector pixel grid. In this particular example, the lenslet array is clocked 26 degrees with respect to the dispersion direction, to avoid overlap of the microspectra. This is best seen in the broadband image. The dispersion direction is aligned with the pixel lines for convenience.

\subsection{IFS propagation}\label{subsec:IFSPropagation}

In order to simulate the IFS optics, the complex wavefront at each of the lenslets should be treated independently, propagated to focus, cropped with a pinhole mask, and further propagated through the collimator, disperser and camera optics (Fig.~\ref{fig:IFS}). This would represent a large amount of computing power and is impractical to use in simulations where hundreds of IFS propagations would be needed.

Instead, our approach is to consider only the total field intensity within a given lenslet, and multiply it by a PSFLet template at the correct location on the detector. This effectively ignores the complex nature of the local electric field, an assumption which is discussed below. Both the PSFLet templates and PSFLet locations can be determined from an optical ray trace software, from lab data (such as PISCES), or by using analytical functions.

We can thus reproduce monochromatic fields by placing the PSFLets at their location and appropriately discretizing them on the detector grid (such as on the left of Fig.~\ref{fig:flatBB}). By summing monochromatic grids of spots at a much finer spectral sampling than the spectrometer's resolution, we can simulate detector maps with smooth broadband input (such as on the right of Fig.~\ref{fig:flatBB}).

%In particular, the PSFLet template is critical to correctly determine electron flux rates and SNR: a more aberrated PSFLet can spread the flux on more detector pixels, hence increasing the overall noise level. The method to obtain PSFLet templates and centroids is discussed extensively elsewhere\cite{Brandt:2017uo}.

\begin{figure}[ht!]
\centering
\includegraphics[width=\textwidth]{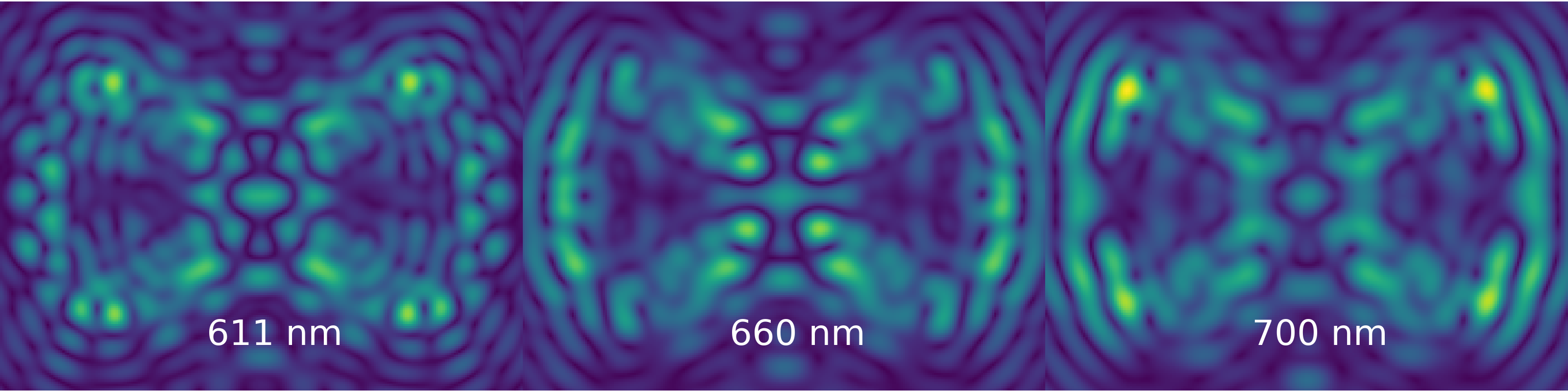}
\caption{Input cube slices into the IFS, showing the coronagraphic PSF for WFIRST's Shaped Pupil bowtie mask for a \SI{1000}{\second} time step. The dark bowtie zones form the ``dark hole", nominally from $\sim 3-9\lambda/D$ from the center of image. These models were generated by JPL's simulation pipeline for the ``OS5" observing scenario (see Section~\ref{subsec:CGIObservingScenarios}). The three different panels show different wavelength slices, illustrating the change in the speckle pattern with wavelength. In the case of this example, the input cube has a total of 45 wavelength slices.}\label{fig:InputSlices}
\end{figure}

\begin{figure}[ht!]
\centering
\includegraphics[width=0.7\textwidth]{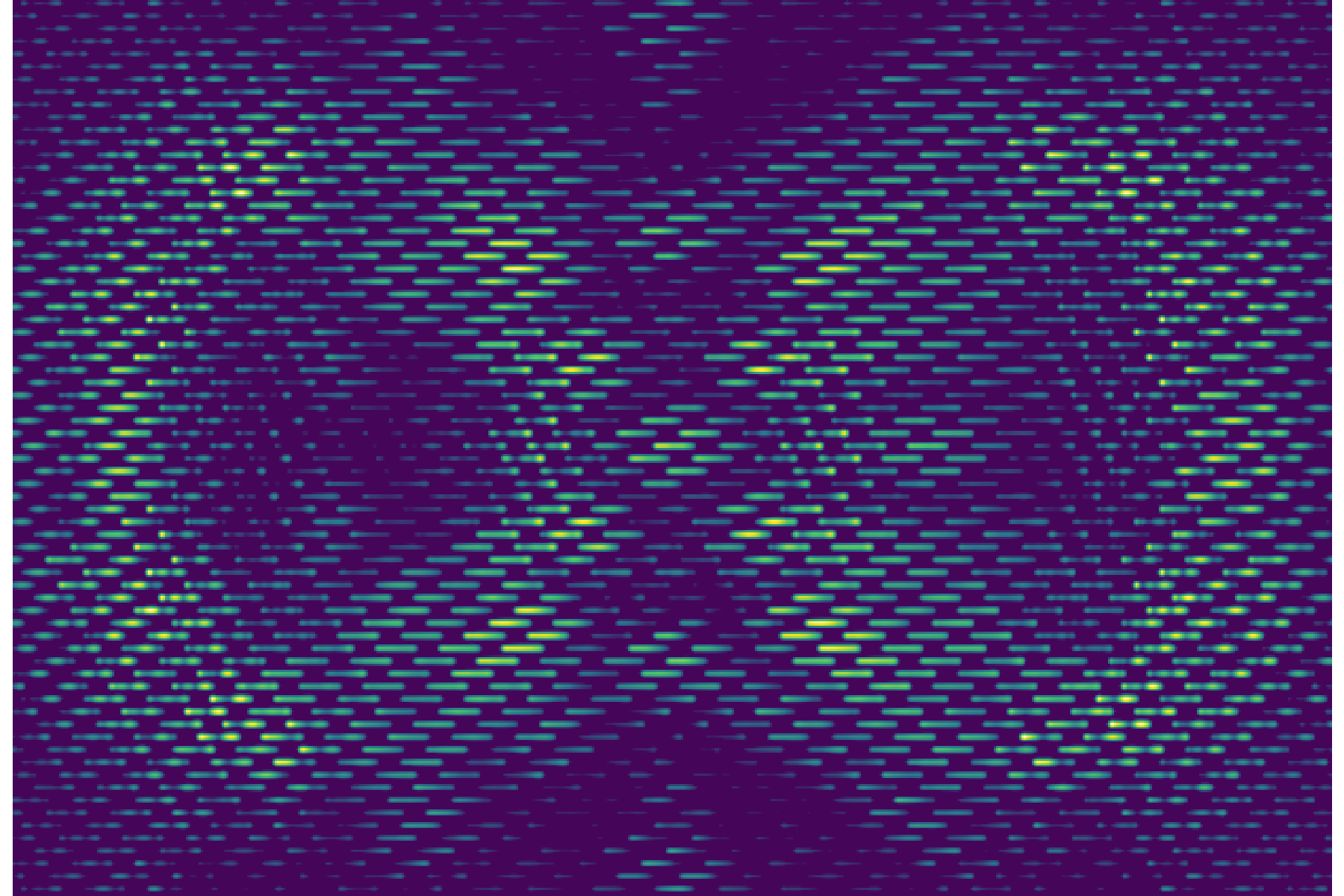}
\caption{IFS detector map corresponding to the input cubes from Fig.~\ref{fig:InputSlices}. The cube is rotated, rebinned onto the lenslet array slice-by-slice, and the flux within each lenslet multiplies a lenslet PSF which location on the detector depends on the wavelength. For a broadband spectrum, each lenslet thus has an elongated microspectrum. For spatial and spectral Nyquist sampling, the information in this image is the same as the one in the input cube, and the input cube can be reconstructed. This detector map corresponds to a simulation of an observing scenario targeting 47 Uma c, which is visible in the right-hand side of the bowtie, close to the inner working angle. We use the same example throughout the paper, and discuss the details in Section~\ref{subsec:CGIObservingScenarios}.}\label{fig:DetectorMap}
\end{figure}

To achieve this, \crispy's propagator requires input datacubes representing the field at the lenslet array (e.g. see 3 slices out of a 45-slices input cube, Fig.~\ref{fig:InputSlices}). If the absolute fluxes are of importance to the user, the input cubes should be in units of photon flux density: each pixel represents the number of photons per second per nanometer of bandwidth within that pixel. The software calculates the appropriate bandwidth from the number of cube slices, assuming the provided list of wavelengths correspond to the band centers for each cube slice. Each slice is treated separately; \crispy{} will rotate and bin the input pixel map onto the lenslet array using the input cube pixel scale, while ensuring flux conservation. The photon flux entering each lenslet, integrated over the short band from a given slice, is used to multiply the corresponding lenslet PSFLet, placed at the correct location on the detector (see the corresponding detector map in Fig.~\ref{fig:DetectorMap}). Because each detector pixel can see photons from multiple wavelengths, all wavelength-dependent effects such as quantum efficiency (QE) and optical efficiency should be included within the input cube itself. Once the detector map is constructed, it is not possible to distinguish between photons from different wavelengths without running an extraction routine.

This provides a fast method to generate IFS cubes, and it can be fully parallelized since each slice is treated separately. The resulting detector maps that \crispy{} generates are now in units of photo-electrons per second. This is the signal that an noiseless detector would read out. 

The processing speed represents a vast advantage over other methods that require propagating each lenslet through all the optical elements individually. A previous version of Goddard's IFS simulation software uses PROPER\cite{Krist:2007gk} to simulate each lenslet, but it was taking $\sim$11~h to run a single simulation on a standard 12-core machine. We have shown excellent agreement between running the full complex propagation for each lenslet, using template PSFLets. The maximum change in the phase gradient across each lenslet is small, even including events required to dig the dark hole, such as DM probes. The resulting wavefront tilt would make the PSFLet coordinates shift by values much smaller than 1 detector pixel. On CHARIS, a lenslet array-based IFS at the Subaru telescope, poor atmospheric Strehl ratio can cause phase tilts across lenslets that can be considerably larger than the ones we expect on WFIRST; nevertheless, the PSFLet map is very stable from one exposure to the next and individual PSFLets do not appear to be shifting over such timescales. 

%Since the lenslet array samples the incoming field at the Nyquist spatial frequency, the local phase within a single lenslet is well-behaved and its tilt, which could change the spot location on the CCD, would only cause a fraction of a pixel shift \mycomment{(BE QUANTITATIVE)} for expected tilt values \mycomment{(CITE Tyler's paper)}. This is consistent with ground-based IFS experience\cite{Brandt:2017uo}: despite large phase variations due to atmospheric perturbations, the PSFLets are not known to move dynamically to within any detectable amount on the IFS final detector \mycomment{(Tyler should check this)}.

%Our general approach then is valid unless the phase variation within individual lenslets becomes so large that it would change the location or shape of the PSFLet, neither of which is expected to happen\mycomment{CHECK THIS?}.

%In order to construct smooth microspectra, it is recommended to oversample the input spectral datacube at least by a factor of 2 compared to the IFS spectral resolving power. This avoids showing ``gaps" in the microspectra. Note that all wavelength-dependent effects to the incoming beam such as the detector response and optical losses should be applied before running the IFS propagation, since now it is not possible to distinguish between photons from different wavelengths until we do a cube extraction.

\subsection{Stochastic EMCCD model}\label{subsec:StochasticEMCCDModel}

WFIRST has baselined an e2v Electron-Multiplying CCD (EMCCD) for its flight operations. This type of detectors can be operated in photon-counting mode, which can get rid of the read noise and offers overall signal-to-noise ratio (SNR) improvements. In order to model the EMCCD, we use a step-by-step stochastic approach\cite{Basden:2003hua,Hirsch:2013iu,Plakhotnik:2006ht,Daigle:2008gu,Harding:2016bc}. First, we multiply the exposure time $t$ by the photo-electron arrival rate map $p$ obtained with \crispy, which takes into account wavelength-dependent quantum efficiency and optical efficiencies. We add a dark current $d$ term and a clock-induced charge (CIC) term $c$, and apply a Poisson distribution to this map $pt+dt+c$. We feed the obtained map of integer electrons to a gain register model, which uses the gamma function to represent the electron multiplying stage\cite{Daigle:2008gu,Hirsch:2013iu}. We add an arbitrary bias $b$ and a zero-mean gaussian process of standard deviation $\sigma$ to describe the read noise  of the electron-multiplying register. We finally apply a threshold at $b+\tau$ (typically $\tau >5.5\sigma$). This allows us to count individual electrons and eliminate the gain register read noise. This model is approximate as it does not take into account effects such as spurious charge that can occur in the EM stage itself\cite{Hirsch:2013iu}. This algorithm is implemented using Python's \Verb|numpy.random| functions in the \crispy{} source code. The set of parameters that we use for the flight simulations is shown in Table~\ref{tab:detectorparams}. All the components of the noise model can be turned off in \crispy's master parameter file.

\begin{table}
\caption[]{Suggested detector parameter settings.}
\begin{center}
\begin{tabular}{l|c|c|l}
Parameter &
Symbol &
Value &
Units \\
\hline
\hline
   
Clock-induced charge & $c$ & 0.01 & \si{\elec\per\pixel\per\frame} \\
Dark current & $d$& \num{2e-4} & \si{\elec\per\pixel\per\second} \\
Electron-multiplying gain & $g$& 2500 & N/A\\
Readout bias & $b$& 200 & \si{\elec\per\pixel\per\frame}\\
Read noise & $\sigma$ &100 & \si{\elec\per\pixel\per\frame}\\ 
Threshold & $\tau$& $5.5\sigma$ & N/A
\label{tab:detectorparams}
\end{tabular}
\end{center}
\end{table}

In order to retrieve the actual photometry in photon-counting mode, we have to account for several effects. The first is to subtract from the average frame a local background level which is sampled in regions of the image with no input flux. The second is to apply a photometric correction caused by coincidence losses at the input of the EM register, which can occur when more than 1 photon is converted into electrons in a single frame. This imposes the frame to be low flux (typically less than 0.1 electron/pixel/frame), in order to keep the coincidence losses from impacting the SNR, which in turns puts constraints on exposure times. A photometry correction for this effect exists\cite{Basden:2003hua} and is $I_{\textrm{corrected}} = -\ln(1-I_{\textrm{original}})$, where $I_{\textrm{original}}$ and $I_{\textrm{corrected}}$ are the images before and after correction, respectively.

\begin{figure}[ht!]
\centering
\includegraphics[width=\textwidth]{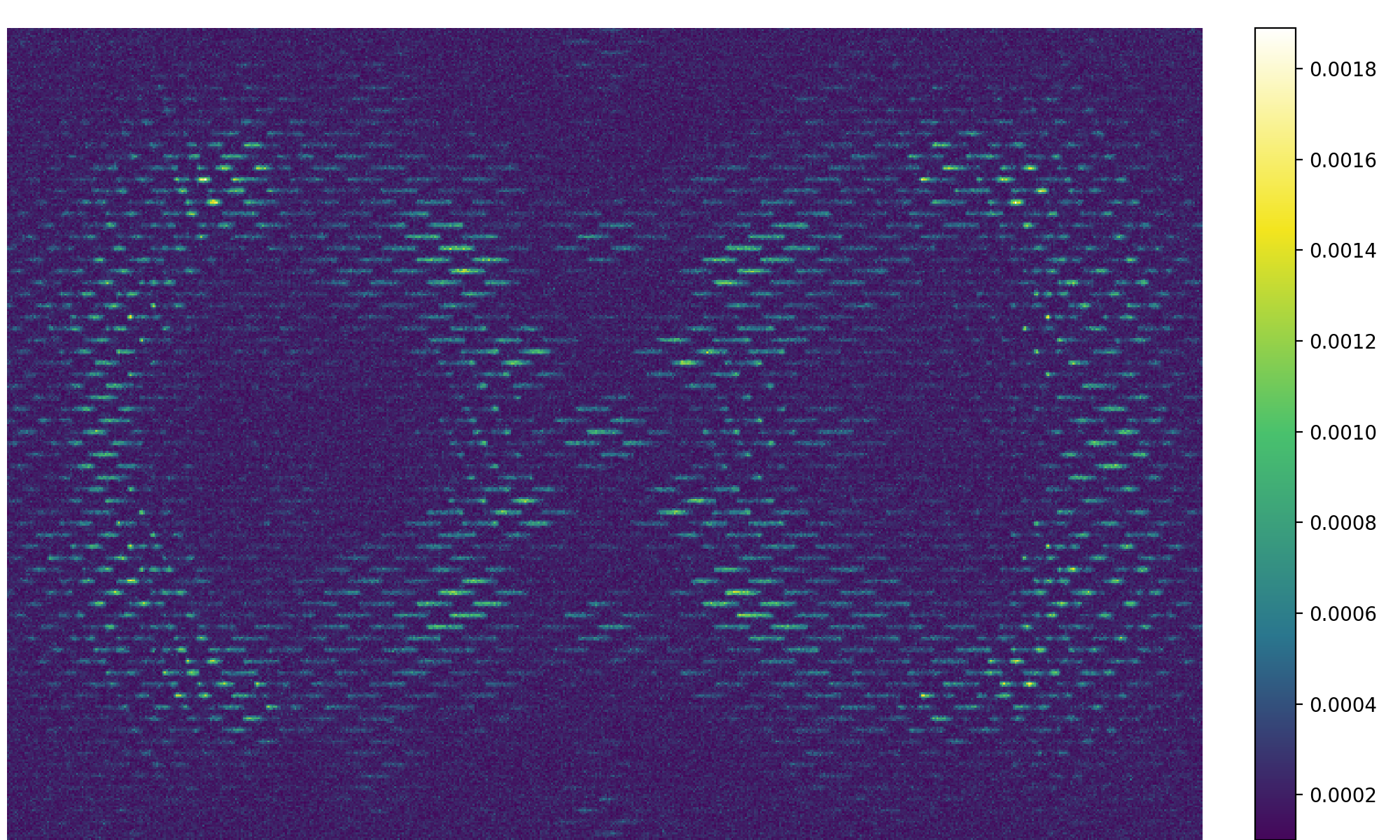}
\caption{Average IFS detector map in electrons per second corresponding to a 27~hr integration time. This average is determined from 1000 individual 100~s exposures, in photon counting mode. The photometry has been corrected to reflect real electron rates.}\label{fig:NoisyDetectorMap}
\end{figure}

A third and final photometric correction occurs, as there is non-zero chance that a photon event might be lower than the threshold applied during photon-counting mode. The likelihood of this happening is $\exp(-\tau/g)$, where $g$ is the EM gain, so we can divide the image by this amount. An example of a noisy map average for a $\sim\SI{27}{\hour}$ observation is shown in Fig.~\ref{fig:NoisyDetectorMap}.

The EMCCD suffers from radiation sensitivity in the space environment\cite{Harding:2016bc}. High energy particles can damage the silicon lattice and create ``traps" which can capture and release electrons at random times during charge transfers to the gain register. As a result, single traps in the regular CCD sensitive area can affect electrons on the entire column, while charge traps that are located in the serial register can affect whole rows. This reduces the effective QE and generates random electron events. Unfortunately, accurately modeling the effects of charge traps is complicated, and also computationally expensive\cite{Nemati:2016bq}. The process cannot be easily parallelized since each new frame requires knowledge of the state of all the traps from the previous frame.

At this moment, \crispy{} is not modeling all of the effects of traps, although the authors are working to increase the speed of existing models\cite{Nemati:2016bq}. At present, we simplify the problem of traps into an overall reduction in QE, using an empirical model developed by JPL during EMCCD testing. This overall QE reduction is a function of the fraction of the mission time the detector spent in the space environment, where a fraction of 1 corresponds to a 6-year mission at the L2 Lagrange point. This model is applied to the average photo-electron rate $p$ before adding the dark current and CIC contributions.

% \begin{lstlisting}
% import numpy as np
% from numpy.random import poisson,gamma,normal

% # phmap is the photon arrival rate map in photo-electron per second (2D array)
% # inttime is the integration time (scalar)
% # dark is the dark current in electrons per second per pixel (scalar)
% # CIC is the clock induced charge in electron per pixel per frame (scalar)

% # This computes the average electron number per frame in the CCD
% average = phmap * inttime + dark * inttime + CIC

% # Generate integer electrons
% atEMRegister = poisson(average)

% # Select only pixels that are non-zero
% EMmask = atEMRegister > 0

% # Create array of zeros
% afterEMRegister = np.zeros(atEMRegister.shape)

% # Apply gain register statistics, EMGain is a scalar representing 
% # the electron-multiplying gain
% afterEMRegister[EMmask] = gamma(atEMRegister[EMmask],EMGain,atEMRegister[EMmask].shape)

% # Read out the gain register, readNoiseSTD and readoutBias are scalar 
% # and expressed in electrons per pixel per readout
% afterReadout = afterEMRegister + normal(readoutBias,readNoiseSTD,afterEMRegister.shape)

% # Photon counting mode. threshold is a scalar in terms of number
% # of standard deviations
% PCmask = afterReadout > readoutBias + threshold*readNoiseSTD

% \end{lstlisting}

%Bijan, and other EMCCD papers; traps in progress, perhaps show image from Bijan's paper ?! We will see how far ahead Ashley gets.

\subsection{IFS extraction}\label{subsec:IFSExtraction}

\begin{figure}[ht!]
\centering
\includegraphics[width=\textwidth]{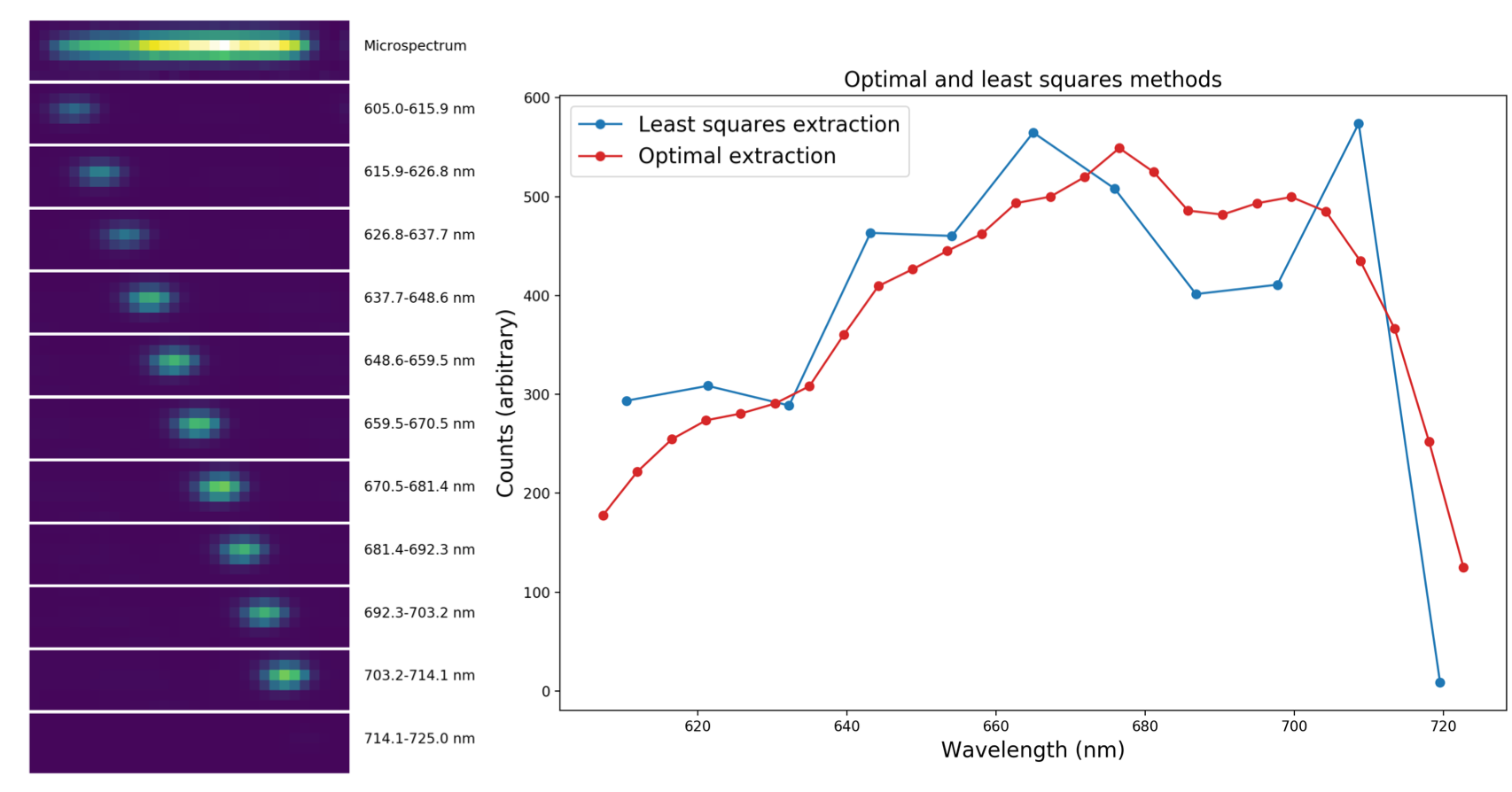}
\caption{Left: A microspectrum (top) is fitted to a weighted sum of template PSFLets in the least squares extraction method, here shown individually. Each PSFLet template is slightly broadened along the dispersion direction to account for finite-width bins in the recovered spectrum. Right: Comparison of the same microspectrum extracted using the least squares method (blue) and the optimal extraction method (red). In the latter, one-dimensional Gaussian functions are fitted along each pixel column of the microspectrum. In this particular configuration (PISCES), the optimal extraction has 26 bins when the least squares extraction only has 11. Note that the bins in the optimal extraction spectrum are not uncorrelated, unlike in the least squares spectrum, which is why the optimal extraction spectrum looks smoother and appears to be taken at a lower resolving power. In order to plot both spectra on top of each other on the same scale, the native optimal extraction spectrum has been scaled by a factor of 1/4. The counts in the recovered spectrum are arbitrary, as each method would have its own specific flux calibration procedure.}\label{fig:OptimalVSlstsqMicrospectrum}
\end{figure}

Once a map of the IFS detector has been generated and the detector readout has been simulated, it remains to extract the original datacube from it. In order to achieve this, \crispy{} extraction routines are heavily based on the work done on the CHARIS data reduction pipeline\cite{Brandt:2017uo}. We include two methods for extraction: one that relies on fitting 1-D Gaussian functions to each pixel across a microspectrum ('optimal' extraction\cite{Horne:1986bg}); and one that considers a least squares fit of the microspectrum as a sum of quasi-monochromatic PSFLet models (least squares extraction\cite{Brandt:2017uo}, see the left-hand side of Fig.~\ref{fig:OptimalVSlstsqMicrospectrum}). The two algorithms are comparable in terms of execution speed, but the least squares extraction requires more calibration data and has a few more parameters. It is also the better of the two algorithms when it comes to producing science-grade reduced datacubes.

Both algorithms rely on knowing precisely the centroids of all the PSFLets for a given spectral bin. This can be achieved using monochromatic flatfields to uniformly illuminate the lenslet array, hence create a quasi uniform grid of spots on the IFS detector (as in the left of Fig.~\ref{fig:flatBB}). The grid can be fitted by low-order 2D polynomials in both the $x$ and $y$ detector coordinates. By cycling through various wavelengths, it is then possible to map exactly where the light is supposed to fall at any wavelength in any microspectrum. 

The least squares extraction method uses a high-resolution model of all the PSFLets across the field. This model can either be an analytical function, a set of templates provided by the user,  or obtained in the lab by dividing up the detector in several regions and deconvolving PSFLets models in each region\cite{Brandt:2017uo}. While the latter is implemented in \crispy{} and its preliminary implementation on the PISCES testbed has proved to be successful, we plan on using detectors with a fine pixel pitch to calibrate the WFIRST IFS on the ground. The deconvolution method is implemented successfully on the CHARIS IFS, but their detector has 4 times the area and 4 times the number of PSFLets, hence improving the PSFLet sampling diversity and as a result, the PSFLet models. One of the main advantage of the least squares extraction method is that it provides residual maps over the entire detector area. These residual maps can be used for calibration purposes, which will be discussed in a later paper.

\crispy{} is also capable of simulating the wavelength calibration step, a critical aspect of WFIRST operations for the IFS. As shown on CHARIS\cite{Groff:2017b}, the PSFLet shapes do not significantly change with time, unlike their location on the detector. As part of their routine observing schedule, single-wavelength updates are used to account for the bulk shift and rotation of the PSFLet grid on the detector. A polynomial fit of the PSFLet grid is done at a single monochromatic wavelength, and the linear and rotational offset from the previous calibration at that wavelength is recorded. These offsets represent the overall translation and rotation of the PSFLet grid as a rigid body, and are thus applied identically to all other wavelengths in the calibration set.

\subsection{Software usage and parameters}\label{subsec:SoftwareParameters}

The publicly released version of \crispy{} is tuned to apply for the WFIRST IFS, but can be applied to any other square lenslet array-based IFS. Hexagonal lenslet geometries are a possible future add-on. The software is open source and publicly available on GitHub\footnote{{\tt https://github.com/mjrfringes/crispy}}. A documentation page is associated with the repository, which has examples and tutorials on how to use the software.

In order to use \crispy{}, the user must choose the basic IFS parameters defined in Table~\ref{tab:crispyparams}. These define the overall geometry of the lenslet array and detector, as well as the spectral properties of the IFS. In addition, the detector parameters mentioned in Section~\ref{subsec:StochasticEMCCDModel} can be populated, with the ability to turn on and off all sources of noise.

%The user has the ability to input its own PSFLet models or use 2D Gaussian PSFLets with specified full-width-at-half-maximum (FWHM). With these parameters, \crispy{} can generate a set of synthetic monochromatic PSFLet grids, just like the ones required for a traditional calibration procedure. %The software is agnostic to the source of the 2D polynomial maps describing the location of the PSFLet grids, as long as it respects a certain formalism. Using the spectral parameters, \crispy{} can generate its own set of synthetic of detector maps with monochromatic flatfields and then run a wavelength calibration procedure on these maps. Alternatively, a wavelength solution constructed from laboratory data can be used instead.

\begin{table}
\caption[]{\crispy{} parameters.}
\begin{center}
\begin{tabular}{l|c|l}
Parameter &
Default &
Description \\
\hline
\hline
\multicolumn{3}{c}{\bf{Lenslet array}} \\
\hline
{\tt nlens} & 108 & Number of lenslets across array\\
{\tt pitch} & \num{174e-6} & Lenslet pitch in meters\\
{\tt interlace} & 2 & Clocking is $\arcsin\left[(\textrm{interlace}^2+1)^{-1/2}\right]$\\
{\tt lensletsampling} & 0.5 & Lenslet size in units of  $\lambda/D$\\
{\tt lensletlam} & 660 & Wavelength at which sampling is defined (in nm)\\
\hline
\multicolumn{3}{c}{\bf{Detector}} \\
\hline
{\tt npix} & 1024 & Number of detector pixels on a side\\
{\tt pixsize} & \num{13e-6} & Pixel size in meters\\
{\tt FWHM} & 2 & Pixels across full-width-at-half-maximum of PSFLet\\
{\tt FWHMlam} & 660 & Wavelength at which the FWHM is defined\\
{\tt gaussian} & True & If True, uses a Gaussian kernel for PSFLets; \\
& & $\qquad$ if False, a kernel needs to be provided\\
\hline
\multicolumn{3}{c}{\bf{Spectrograph}} \\
\hline
{\tt BW} & 0.18 & Fractional bandwidth\\
{\tt R} & 50 & Spectral resolving power\\
{\tt npixperdlam} & 2 & Pixel sampling of individual spectral channel\\
{\tt nchanperspec\_lstsq} & 1 & Number of channels per native spectral element

\label{tab:crispyparams}
\end{tabular}
\end{center}
\end{table}

Without any a priori knowledge other than those parameters, \crispy{} generates a set of monochromatic flatfields at several user-defined wavelengths in the band. The nominal centroids for {\tt FWHMlam} are determined using an ideal polynomial function with no geometric aberration. The users are also able to manually input their own distortions (e.g. to simulate a distortion that was measured on a testbed).
The spots are 2D Gaussian PSFLets ({\tt gaussian} set to {\tt True}). It assumes the resolving power {\tt R} is constant across the spectrum and the detector samples a spectral element with {\tt npixperdlam} pixels for the central wavelength {\tt FWHMlam}. The deviation in pixels along the dispersion direction from the centroid at wavelength {\tt FWHMlam} is set by the following equation:
\begin{equation}
\Delta\textrm{pix}(\lambda) = \tt{npixperdlam}\times\tt{R}\times\ln\frac{\lambda}{\tt{FWHMlam}}.
\end{equation}

\crispy{} then runs a wavelength calibration routine on the set of PSFLet maps. This produces an array of polynomial coefficients for each wavelength, which fully characterize the PSFLet spot locations in the 2D detector coordinates. \crispy{} generates all the required calibration products which are very similar to the ones used by the CHARIS data reduction pipeline. Importantly, it is possible to use noisy images for this step, which allows us to better understand what is the impact of errors in the wavelength calibration phase on final data products. During this phase, the user is allowed to tweak the spectral resolution to produce more or less slices in the extracted cube.

Once the calibration step is finished, \crispy{} can  process input cubes into IFS detector maps, and can extract these detector maps into spectral datacubes. Input cubes need to provide certain FITS header keywords such as the wavelength and the pixel scale, which \crispy{} will use to rotate and rebin the input cube to match the lenslet array geometry. The IFS map is then generated by combining monochromatic maps for each wavelength of the input cube.

The detector noise parameters will determine the SNR of the extracted cube. Although \crispy{} uses a EMCCD noise model, individual noise contributors can be turned off in order to mimic a regular CCD as well. In the case where an EMCCD model is chosen, the user needs to pay attention to the electron flux at the final detector map which should not exceed $\sim$\SI{0.1}{\elec\per\pixel\per\frame} if one wants to avoid coincidence losses.

The extraction algorithm can finally be applied to the noisy detector maps. It will use the calibration products that were generated during the wavelength calibration phase. Using both the optimal extraction and the least squares extraction method, a spectral datacube is generated. 

\crispy{} is evolving fast and several improvements are planned that will make it easier for users to input their own distortion maps, their own field-dependent PSFLet templates, and facilitate other operational aspects of the software.

\section{Application to WFIRST CGI-IFS}\label{sec:Application to WFIRST CGI-IFS}

\crispy{} is being developed for WFIRST and the CGI Science Investigation Teams as the primary IFS simulator. Because of its intricacies, it warrants a standalone and focused effort, the development of which has been accelerated by the IFS similarities to existing ground-based IFS instruments such as CHARIS. We have been able to enjoy a productive collaboration with the CHARIS team and were able to apply most of their IFS pipeline for the WFIRST IFS. 

\crispy{} is intended to be used largely in three main ways at this stage of the WFIRST mission. First, it is used routinely on the PISCES testbed for coronagraph closed-loop control. This is discussed extensively in a companion paper in these proceedings\cite{Groff:2017}.

Second, \crispy{} is intended to be used to simulate realistic data products, similar to the ones that will be produced by the flight instruments. This is discussed in Section~\ref{subsec:CGIObservingScenarios}.

Finally, \crispy{} is used by the Goddard IFS team to study and improve the IFS design. As it is capable of simulating science observations and can determine quantities such as SNR, it can be used as a good metric to establish tolerances on the IFS and to explore the design parameter space. This is illustrated in Section~\ref{subsec:Trade-offsAndDiscussion}.

% \subsection{Application to PISCES}\label{subsec:ApplicationToPISCES}

% % \begin{figure}[ht!]
% % \centering
% % \includegraphics[width=\textwidth]{PISCES_PSFLets.png}
% % \caption{ }\label{fig:PISCES_PSFLets}
% % \end{figure}

% From day one \crispy{} was designed to be compatible with the PISCES data products, and can ingest real testbed data just as well as it can simulate its own. Hence, the wavelength calibration methods and extraction routines are constantly being validated by testbed operations through consecutive updates and improvements, in order to spot any compatibility issues early on and ensure that the no software element needed for the closed-loop control is being overlooked. 

% The cube extraction method has been integrated to the closed-loop control system at HCIT. The software is crucial to reconstruct broadband dark hole taken with the IFS and separate the speckle pattern into its different wavelengths. The wavelength slices are then used by the control system, that applies the corrections to the deformable mirrors in order to suppress the speckles. 

% This has allowed to obtain high-contrast ($\sim\num{1e-8}$) dark holes over 18\% bandpass, which is considerably wider than the standard 10\% bandpass control that was achieved by HCIT before. More details about the exciting PISCES results and the future steps are proposed in a companion paper in these proceedings\mycomment{cite tyler's paper}.

\subsection{CGI observing scenarios}\label{subsec:CGIObservingScenarios}

% This is intended both as a way to validate parametric sensitivity and SNR estimates and to develop and assess post-processing algorithms. The latter is an important part of the design of a coronagraph, as the detectability of a faint planet often relies on our ability to suppress time-varying speckle patterns inside the dark hole. While ground-based IFS can routinely achieve post-processing contrast gains of 10 or 20, it is not clear that it will be the same for the WFIRST IFS since the contrast levels will be two or three orders of magnitude deeper than what can be achieved from the ground.

Realistic data products that mimic an observing scenario are helpful to prepare post-processing pipelines, assess observing strategies, and, in the particular case of a coronagraph, understand how much can be gained by post-processing, which can relax the requirements on the instrument's raw contrast goals.

%\crispy{} can process realistic coronagraph simulations outputs, which exhibit the most up-to-date CGI behavior when the low-order wavefront control is turned on. Importantly, these simulations can take into account the noise injected by the WFIRST telescope and control system, which is one of the critical modeling efforts at this stage of the mission.

\begin{figure}[H]
\centering
\includegraphics[width=0.8\textwidth]{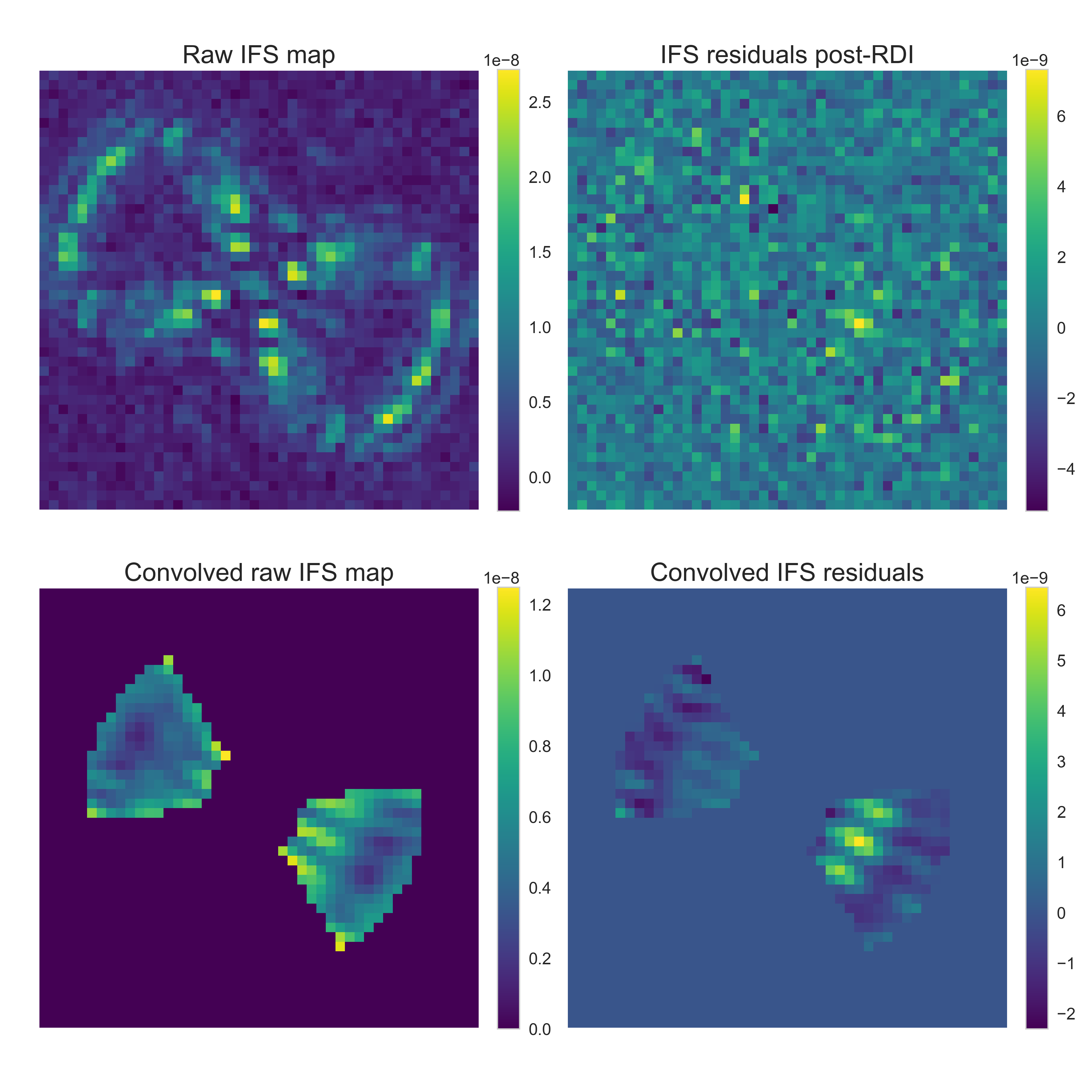}\label{fig:RDImaps}
\caption{Sample of \protect\crispy{} products for an observing scenario. For each of the four panels, only the middle slice (660~nm) is shown. The top left shows the raw IFS map averaged over 1000 individual exposures of 100~s. The planet is barely visible on the right half of the bowtie, close to the inner working angle. The top right corresponds to the residuals after RDI, which uses a nearby reference star to measure the speckle pattern. The bottom left shows the top IFS map convolved with a matched filter that was constructed using an ideal off-axis PSF. The planet is now more visible in this image. The bottom right shows the convolved residuals. At this point the planet is clearly visible many standard deviations above the local noise.}
\end{figure}

\crispy{} can ingest a realistic WFIRST CGI observing scenario time series of spectrally and spatially resolved cubes. The fiducial WFIRST scenario mimics an observation of the giant planet 47~Uma~c that was discovered through radial velocity techniques, using $\beta$~Uma as a close-by reference star. This scenario is labeled ``OS5" in the rest of this section. It divides the integration time into chunks of 1000~s (subsequently 1~ks), during which we assume that the speckle field does not change. 30~ks are spent on the reference star, $\beta$~Uma, and 100~ks are spent on the target star, 47~Uma. Importantly the simulation includes realistic noise expected from the observatory, as well as the low-order wavefront control response. 

The module propagates all of the individual cubes into IFS maps representing the photon arrival rate at each discrete time step, both for the reference and target stars. For each of these maps, we select the correct exposure time that ensures less than 0.1~electron/frame in most areas of the frame. Using this exposure time, typically about 10~s for the reference and 100~s for the target, \crispy{} computes as many noise realizations as needed to amount to the total exposure time for this time step. For example, if each time step corresponds to 1000~s, then 10 frames of 100~s are simulated per step for the target. All of the frames from all time steps are averaged separately for the reference and target, background-subtracted using non-illuminated parts of the detector, and the photometry is corrected as discussed in Section~\ref{subsec:StochasticEMCCDModel}.

Once a noisy average is determined for both the reference and the target star, both maps are extracted into noisy IFS cubes. We implement a reference differential imaging technique (RDI) to subtract  from the target cube the linear combination of the reference cube that minimizes the least squares residuals within the dark hole. Each slice is treated independently. The residual cube is then convolved by a matched filter cube and normalized to contrast units using an ideal stellar off-axis PSF. The planet's spectrum corresponds to the spectrum of the pixel which matches the planet's location, assumed known in this scenario.

Several by-products of this module are shown in Fig.~\ref{fig:RDImaps}, where one slice of a noisy simulation is shown. In this case, the planet is at the brightest location of the convolved IFS residuals (bottom right panel), and its contrast is the maximum pixel value.

\begin{figure}[H]
\centering
\includegraphics[width=0.9\textwidth]{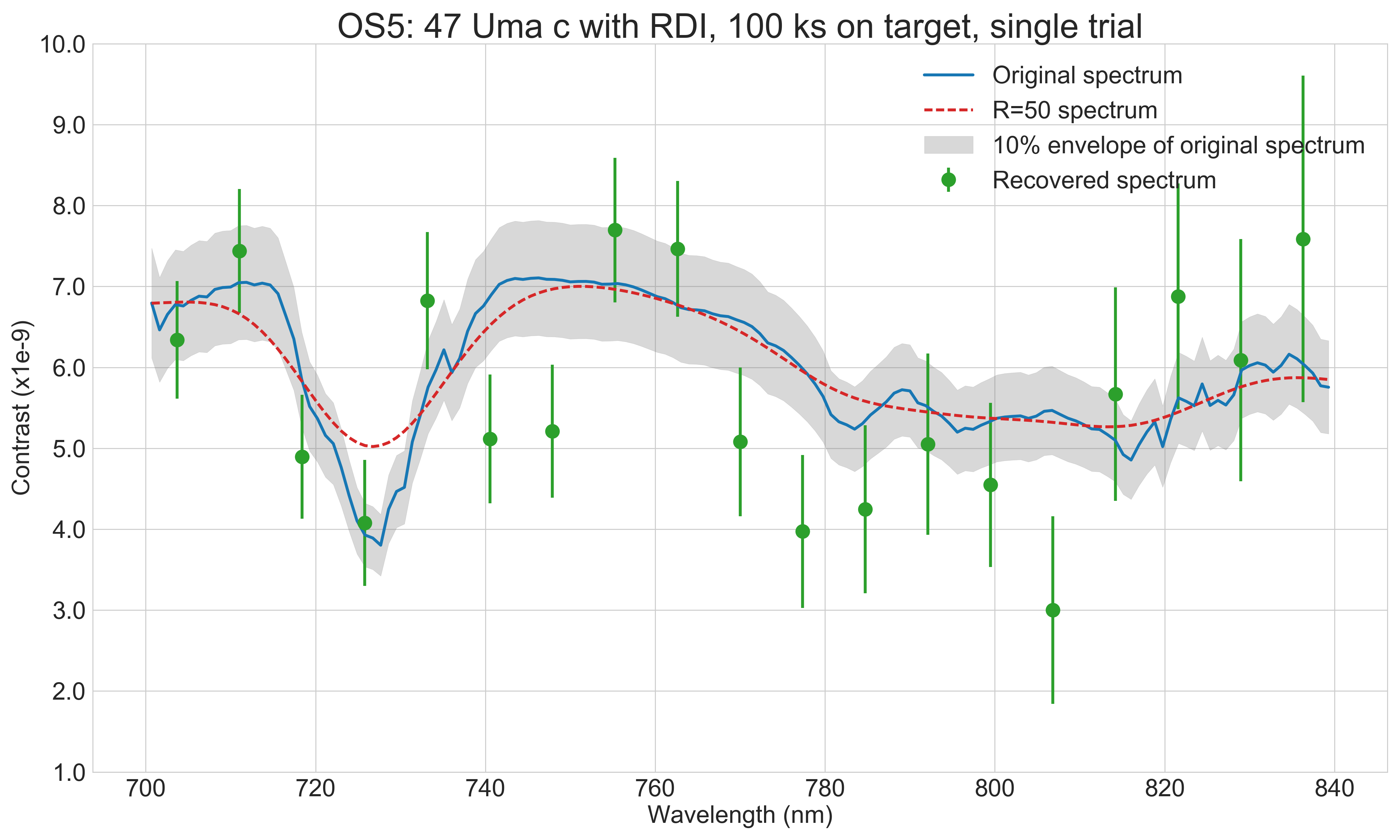}\label{fig:SNR770}
\caption{Extracted spectrum\cite{Cahoy:2010gz} from an OS5 trial of 27~h exposure time. The recovered spectrum is shown along with the original and convolved spectrum. The gray area indicates the SNR=10 confidence envelope of the original spectrum for visual reference. If the SNR was 10, the error bars would be the size of this envelope.}\label{fig:SNRs}
\end{figure}
While the SNR in the bottom right panel of Fig.~\ref{fig:RDImaps} appears large, it is not representative of the actual measurement uncertainty. Indeed, when running multiple realizations of the entire OS5 scenario, the variation in the contrast values that we obtained post-RDI varied significantly more than the SNR within a single realization would predict. As a result, we conclude that the SNR within a single OS5 trial is not capturing the uncertainty correctly: for example it is not capturing photon noise from the source. 
In order to determine an accurate SNR and put correct error bars on the recovered spectra, we choose to run our 100 OS5 trials, and use the standard deviation in the determined contrast for each slice as our error bar. A recovered spectrum\cite{Cahoy:2010gz} for the OS5 scenario is shown in Fig.~\ref{fig:SNRs} with representative error bars. The SNR obtained with our method agrees with parametric SNR estimates for the CGI-IFS to within 20\%.

\subsection{Trade-offs and discussion\label{subsec:Trade-offsAndDiscussion}}

An IFS simulator like \crispy{} is a good tool to explore design trades, as it allows the use of accurate metrics to assess performance. The sensitivity of the chosen metric to various design parameters can be assessed and used to establish tolerances on the instrument. The simulator also provides a useful guide to adjust the design in order to maximize the chosen metric. 

As an example, we show here the result of a simulation concerning the lenslet sampling at the lenslet array. It had been assumed that the lenslet array would sample the incoming beam at the Nyquist sampling ($0.5\lambda/D$) for the shortest wavelength in the IFS, \SI{600}{\nano\meter}, which is the same as saying we will have 2.2 lenslets per $\lambda/D$ at \SI{660}{\nano\meter}. This is justified mostly by the need to interpolate the IFS cubes between the reference and target star in order to subtract the reference PSF, in the case that they are offset from each other. The concern is that if we sample coarser than Nyquist, any offset would cause issues with the interpolation and would dramatically reduce the potential post-processing gains that can be achieved. This lenslet sampling parameter has far-reaching implications, since it sets the number of lenslets that will be used to sample a planet's PSF, which then spreads the planet's light onto more detector pixels, which will in turn affect our SNR.

\begin{figure}[H]
\centering
\includegraphics[width=\textwidth]{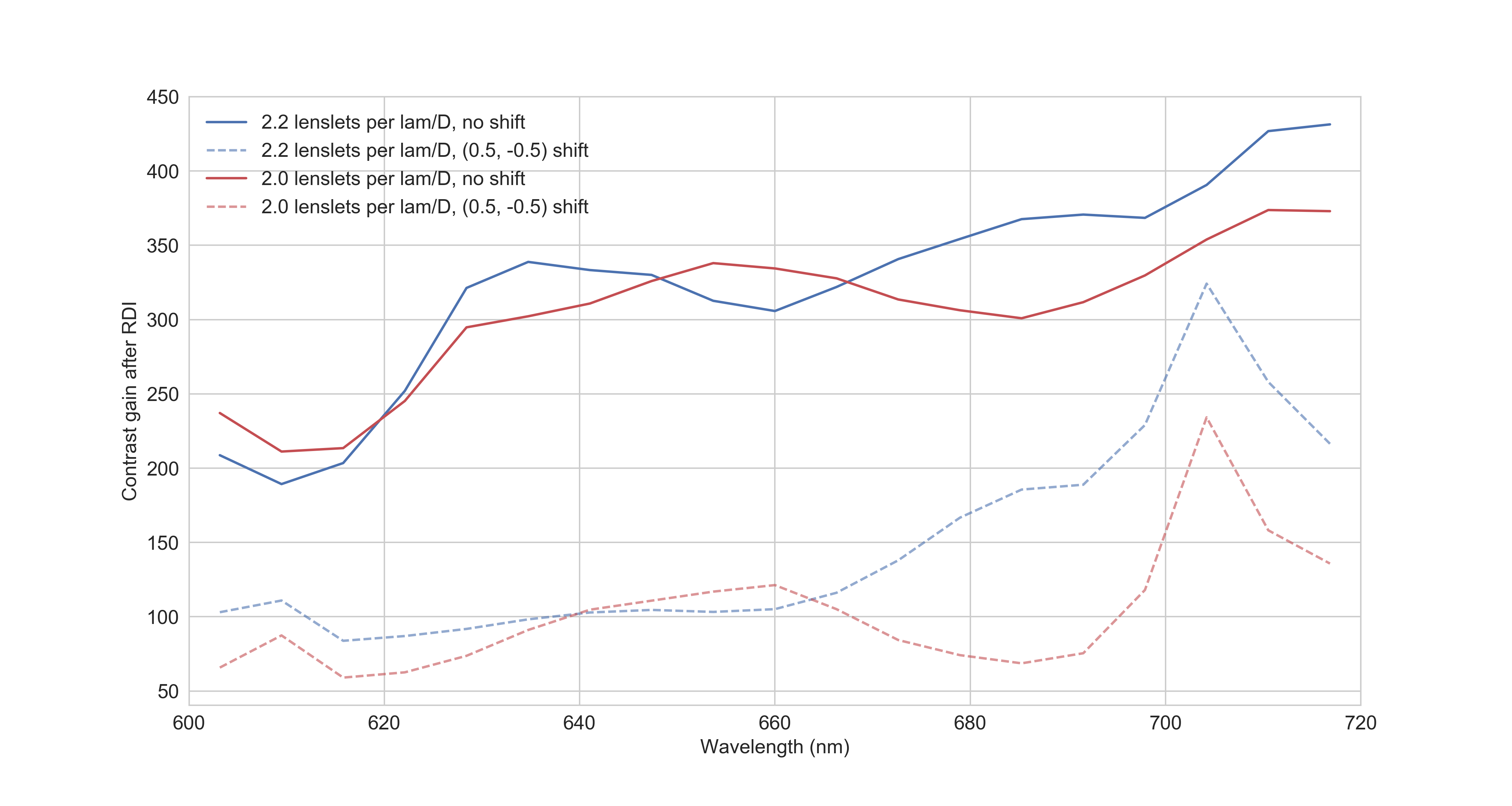}
\caption{Contrast gain curves showing how much the standard deviation of the convolved speckle is improved by our RDI method. This is done for the case where both reference and target stars are aligned, and when the reference star is misaligned with respect to the target star in the worst possible case (0.5 lenslet in both directions). This is plotted for both 2.2 and 2.0 lenslets per $\lambda/D$ at \SI{600}{\nano\meter}. Even with the coarser sampling, the post-processing gain is still larger than 50 even at the shortest wavelength, which indicates that it will likely not be a limitation once real noise is taken into account.}\label{fig:contrastgains}
\end{figure}

\crispy{} allows us to model this entire process and quantify the impact of the lenslet sampling on the post-processing gain. With this example, we are trying to determine whether it is possible to use a bit coarser sampling and not degrade post-processing gain too much. We set up the experiment with a noiseless detector, in order to understand what is the maximum possible post-processing gain that can be achieved. We add a known offset to the reference star with respect to the target star on the lenslet array. This is feasible with very little error since the fine sampling of the input cube from JPL's integrated simulation pipeline is typically given at 10 pixels per $\lambda/D$, significantly more than the Nyquist sampling. Both the offset reference cube and aligned target cube are propagated through the IFS and extracted back using our IFS extraction tool. The IFS cube of the reference star is then counter-shifted by the correct amount (assumed known), but now it has a 2.2 lenslets per $\lambda/D$ at \SI{660}{\nano\meter} (instead of 10 when we did the first interpolation), which will introduce interpolation noise. RDI is applied using the countershifted reference IFS cube, and we measure the ratio of the convolved target star residuals with RDI and the convolved target star without RDI (similar to Fig.~\ref{fig:RDImaps} but without any noise and without any planet in the field of view). We repeat this process for the case of a slightly coarser sampling of 2 lenslets per $\lambda/D$ at \SI{660}{\nano\meter}.

The wavelength-dependent contrast gains in the noiseless case is shown in Fig.~\ref{fig:contrastgains} for both sampling cases. In this figure, \Verb|lam| is equal to \SI{660}{\nano\meter}. At the finer sampling of 2.2 lenslets per $\lambda/D$, the contrast gain is $>200$ for the case where both reference and target are aligned on the lenslet array. In the worst interpolation case possible, which corresponds to an interpolation in each direction of the lenslet array by 0.5 lenslet, the contrast gain drops to $\sim$65 for the coarser sampling and $\sim$100 for the finer sampling. While this represents a loss in contrast gain by a factor of $\sim$4, the remaining contrast gain in the case of the coarser sampling is still large, and will not limit our ability to remove the speckles. In any real-world instrument, our maximum achievable contrast gain will likely not exceed 10 or 20, which is the state-of-the-art improvement achievable by ground-based instruments. 

\begin{figure}[H]
\centering
\includegraphics[width=\textwidth]{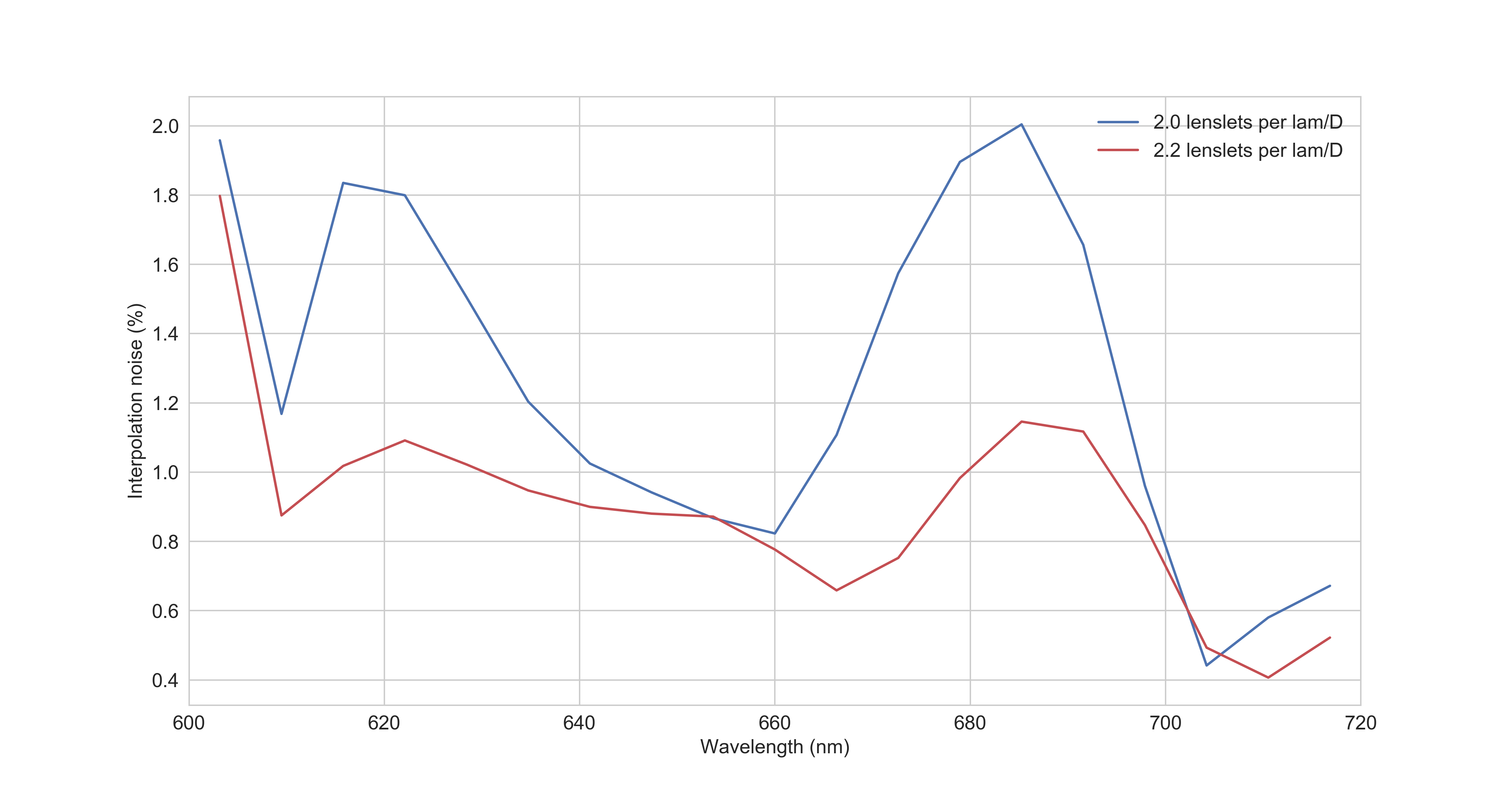}
\caption{Interpolation noise caused by shifting the reference star by half a lenslet in both directions, which corresponds to the worst possible interpolation case. The noise is calculated by computing the RMS of the difference between the cubes when the aligned reference is propagated, and when the reference is shifted, propagated then countershifted. This is a measure of how much difference there is between the two cubes with and without the shift. Dividing by the RMS of the propagated aligned reference allows us to normalize this quantity. Between the two lenslet sampling values, the interpolation noise is increased by less than a factor of 2, but always remains at less than 2\%, which indicate that the maps are very much alike despite the slightly coarser sampling.}\label{fig:interpolationnoise}
\end{figure}

The amount of interpolation noise is assessed by computing the RMS of the difference between the cubes when the aligned reference is propagated, and when the reference is shifted, propagated then countershifted. This is a measure of how much difference there is between the two cubes with and without the shift. We divide by the RMS of the aligned reference cube for normalization (Fig.~\ref{fig:interpolationnoise}).

Thus, this simulation offers strong indication that sampling the lenslet array a 2 lenslets per $\lambda/D$ at $\lambda=\SI{660}{\nano\meter}$ will not limit our ability to use algorithms like RDI. It  will however allow for an overall number of detector pixels per planet PSF reduced by $\sim$20\%, thus improving the SNR of our observation. A more comprehensive study that analyses even coarser sampling is underway to find the optimal parameter.

\section{Conclusions}

\crispy{} is a versatile IFS simulator and extractor for the WFIRST IFS. It can be used to simulate the IFS sensitivity and performance, study trade-offs, generate realistic data products, and is also routinely used for operations on the PISCES testbed at JPL. The software is open source and publicly available on GitHub\footnote{{\tt https://github.com/mjrfringes/crispy}}. 

\crispy{} uses a simplified propagation model to construct IFS images instead of propagating the complex field through each lenslet. It uses a stochastic model to represent the noise and response of electron-multiplying CCDs, and implements photon-counting algorithms for optimal performance. Finally, it is founded on the CHARIS data reduction pipeline to extract IFS maps back into spectral datacubes. 

We used the software to simulate a realistic observing scenario for the WFIRST coronagraph. We implemented a reference differential imaging technique and recovered accurate planet photometry with SNR that is close to the parametric estimates used by the CGI yield calculation team. We also constructed data products that can be used for assessment and improvement of post-processing techniques for the CGI instrument. 
We used \crispy{} to determine that changing the lenslet sampling on the IFS lenslet array would not undermine our efforts to obtain good post-processing gains. This is the first of multiple trade-off studies that we plan to do in the upcoming months and years for WFIRST.

%%%%%%%%%%%%%%%%%%%%%%%%%%%%%%%%%%%%%%%%%%%%%%%%%%%%%%%%%%%%%
\acknowledgments    
 
The authors would like to thank Bijan Nemati for  sharing the WFIRST parametric estimate model and teaching us how to use it. We would like to also thank Leon Harding, Michael Bottom and Patrick Morrissey for helpful discussions about the modeling of the EMCCD. Portions of this research was carried out at the Jet Propulsion Laboratory, California Institute of Technology, under a contract with the National Aeronautics and Space Administration.

%%%%%%%%%%%%%%%%%%%%%%%%%%%%%%%%%%%%%%%%%%%%%%%%%%%%%%%%%%%%%
%%%%% References %%%%%

\bibliography{references}   %>>>> bibliography data in report.bib
\bibliographystyle{spiebib}   %>>>> makes bibtex use spiebib.bst

\end{document}